\documentclass[twoside]{article}
\usepackage[dvips]{graphicx}
\usepackage{qic,epsfig}
\usepackage{bm}
\usepackage{amsmath,amssymb}
\usepackage{type1cm}

\newtheorem{theo}{Theorem}
\renewcommand{\thefootnote}{\fnsymbol{footnote}}  
\DeclareSymbolFont{lettersA}{U}{txmia}{m}{it}
 \DeclareMathSymbol{\Indi}{\mathord}{lettersA}{'211}
 \usepackage{latexsym}
\begin{document}
\setlength{\textheight}{8.0truein}    
          
\thispagestyle{empty}
\setcounter{page}{1}

\vspace*{0.88truein}

\alphfootnote

\fpage{1}
\centerline{\bf The stationary measure for diagonal quantum walk with one defect}
\vspace*{0.37truein}
\centerline{\footnotesize
Takako Endo\footnote{endo.takako@ocha.ac.jp\;(e-mail of the corresponding author)}}
\vspace*{0.015truein}
\centerline{\footnotesize\it Department of Physics, Ochanomizu University}
\baselineskip=10pt
\centerline{\footnotesize\it 2-1-1 Ohtsuka, Bunkyo, Tokyo, 112-0012,
Japan}
\vspace*{10pt}
\centerline{\footnotesize\it LERMA, Observatoire de Paris, PSL Research
University, CNRS,
Sorbonne Universit$\acute{e}$s, UPMC Univ. Paris 6, UMR 8112, F-75014}
\baselineskip=10pt
\centerline{\footnotesize\it 4 place Jussieu 75005 Paris, France}
\vspace*{10pt}
\centerline{\footnotesize 
Hikari Kawai\footnote{kawai-hikari-dy@ynu.jp}}
\vspace*{0.015truein}
\centerline{\footnotesize\it Department of Appllied Mathematics, Faculty of Engineering, Yokohama National University}
\vspace*{10pt}
\centerline{\footnotesize 
Norio Konno\footnote{konno@ynu.ac.jp}}
\vspace*{0.015truein}
\centerline{\footnotesize\it Department of Appllied Mathematics, Faculty of Engineering, Yokohama National University}
\baselineskip=10pt
\centerline{\footnotesize\it 79-5 Tokiwadai, Hodogaya, Yokohama, 240-8501, Japan}
\vspace*{10pt}

\vspace*{0.225truein}
\vspace*{0.21truein}

\begin{abstract}This study is motivated by the previous work \cite{takei}. We treat 3 types of the one-dimensional quantum walks (QWs), whose time evolutions are described by diagonal unitary matrix, and diagonal unitary matrices with one defect.  
In this paper, we call the QW defined by diagonal unitary matrices, ``the diagonal QW'', and we consider the stationary distributions of generally 2-state diagonal QW with one defect, 3-state space-homogeneous diagonal QW, and 3-state diagonal QW with one defect. 
One of the purposes of our study is to characterize the QWs by the stationary measure, which may lead to answer the basic and natural question, `` What the stationary measure is for one-dimensional QWs ? ''.
In order to analyze the stationary distribution, we focus on the corresponding eigenvalue problems and the definition of the stationary measure.\end{abstract}

\section{Introduction}
\label{Introduction}
Quantum walk (QW) is a quantum mechanical version of classical random walk, whose time evolution of QWs are defined by unitary evolutions of probability amplitudes. For its characteristic properties, QW has attracted much interest of various fields in the quantum scale, such as quantum algorithms \cite{ambainissan, shenvi} and quantum physics \cite{gos, kitagawa, oka, wojcik}.
Many reviews and books on QWs have been published so far \cite{cantero, kempe, konnoqw, por, venegassan, venegas}.
Owing to the wealth applications, it is beneficial to study QWs both theoretically and experimentally. 
In recent years, QWs have also been implemented experimentally by several kinds of materials, such as trapped ions \cite{zari} and photons \cite{peret}. 
However, because of its difficulties for QWs to implement the state after many steps, we have not succeeded to see experimentally the behavior in the long time-limit. In addition, for the quantum nature, it is hard for us to intuitively assimilate the properties. \\
\indent
Recently, to derive measures from QWs has become one of the hottest topics in the theoretical study of QWs \cite{watanabe,watanabesan,endo,endosan,uniform,takei}. Especially, the properties of the stationary measures of the two-state  QWs in one dimension have been gradually studied in the immediate past \cite{watanabe,watanabesan,endosan,uniform,takei}. 
As is well known, the stationary measure of the Markov chain has been deeply studied, however, the fruits of study for QW has not yet been given.
This is one of the motivations of our study.
In this paper, we focus on the 2-state and 3-state QWs in one dimension whose time evolutions are generally expressed by a diagonal unitary matrix or diagonal unitary matrices with one defect.\\
\indent
Now we explain concisely the previous studies of stationary measures. In $2013$, Konno et al. \cite{segawa} gave a stationary measure of the QW with one defect whose quantum coin has a phase at $x=0$, and Konno \cite{uniform} gave the uniform measure as a stationary measure of the one-dimensional QWs in one dimension.
Then Endo et al. \cite{watanabesan} got a stationary measure of the QW with one defect whose quantum coins are defined by the Hadamard matrix at $x\neq0$ and the rotation matrix at $x=0$.
Endo and Konno \cite{watanabe} derived a stationary measure of ``the Wojcik model'' in $2014$, and then Endo et al. \cite{endosan} obtained a stationary measure of the two-phase QW. 
Konno and Takei \cite{takei} showed that the set of the stationary measures of the QW except $U$ is diagonal, contains non-uniform stationary measure and they showed that any
stationary measure is uniform for diagonal matrices. Moreover, they proved that the set of the stationary
measures contains uniform measure for the QW in general. 

The organization of this paper is as follows. In the next section, we give a description of $n$-state discrete-time QW (DTQW) in one dimension. We consider the stationary measures of the 2-state diagonal QW with one defect in Section \ref{onedefectqw}, and then, we study 3-state case in Section \ref{3state-1defect}. 


\section{Description of $n$-state DTQW in one dimension}

As a quantum version of classical random walk with discrete-time, discrete-time QW (DTQW) has been intensively studied since the beginning of 2000'. In this section, we explain briefly how the $n$-state DTQW is generally described, where $n\in\mathbb{N}$.
The total space is defined by a Hilbert space ${\cal H}$:
\[{\cal H}={\cal H}_{P}\otimes{\cal H}_{C},\]
where
${\cal H}_{P}$ is spanned by $\{|x\rangle;x\in\mathbb{Z}\}$, called a position Hilbert space, and ${\cal H}_{C}$ is a coin Hilbert space, spanned by $\{|J\rangle;J\in\{L,R\}\},$ where $\mathbb{Z}$ is the set of integers.
The time evolution of the walk is described by a set of $n\times n$ unitary matrices $\{U_{x}\}_{x\in
\mathbb{Z}}$ on ${\cal H}_{C}$, where
\[U_{x}=\begin{bmatrix}
x_{11} & x_{12} &\cdots & x_{1n} \\
x_{21}& \ddots&&\vdots \\
\vdots&&\ddots&\vdots\\
x_{n1}&\cdots&\cdots&x_{nn}
\end{bmatrix}\quad(x_{ij}\in\mathbb{C},i,j\in\mathbb{N}),\]
which is called {\it the quantum coin}.
Note that the subscript $x\in \mathbb{Z}$ represents the position of the walker. 
To define the time evolution, let us divide the matrix $U_{x}$ into $n$ parts as follows:
\[U_{x}=V_{1}+V_{2}+\cdots+V_{n},\]
where
\[V_{1}=\begin{bmatrix}
x_{11} & x_{12} &\cdots & x_{1n} \\
0 & \ddots&&0\\
\vdots&&\ddots&\vdots\\
0 & \cdots&\cdots&0\\
\end{bmatrix},\quad  V_{2}=\begin{bmatrix}
0 & \cdots&\cdots&0\\
x_{21} & x_{22} &\cdots & x_{2n} \\
\vdots&&\ddots&\vdots\\
0 & \cdots&\cdots&0\\
\end{bmatrix},\cdots, V_{n}=\begin{bmatrix}
0 & \cdots&\cdots&0\\
\vdots&\ddots&&\vdots\\
0 & &\ddots&0\\
x_{n1} & x_{n2} &\cdots & x_{nn} \\
\end{bmatrix}
.\]
As the general setting of DTQW, the inner state of the walker consists of $n$ chiralities:
\[|W_{1}\rangle= {}^T\![1, 0, \cdots, 0]_{n}, |W_{2}\rangle= {}^T\![0, 1, \cdots, 0]_{n},\cdots,|W_{n}\rangle={}^T\![0, 0,\cdots, 1]_{n}.\]
At each time step, the walker steps according to the chirality as follows:
\begin{enumerate}
\item Case of $|W_{1}\rangle$: 
\begin{itemize}
\item if $n$ is odd, then the walker moves to the left with $(n-1)/2$ steps.
\item if $n$ is even, then the walker moves to the left with $n/2$ steps.
\end{itemize}
\item Case of $|W_{2}\rangle$: 
\begin{itemize}
\item if $n$ is odd, then the walker moves to the left with $(n-3)/2$ steps.
\item if $n$ is even, then the walker moves to the left with $(n-2)/2$ steps.
\end{itemize}
\item Case of $|W_{n-1}\rangle$: 
\begin{itemize}
\item if $n$ is odd, then the walker moves to the right with $(n-3)/2$ steps.
\item if $n$ is even, then the walker moves to the right with $(n-2)/2$ steps.
\end{itemize}
\item Case of $|W_{n}\rangle$: 
\begin{itemize}
\item if $n$ is odd, then the walker moves to the right with $(n-1)/2$ steps.
\item if $n$ is even, then the walker moves to the right with $n/2$ steps.
\end{itemize}
\end{enumerate}   
Furthermore, if $n$ is odd and the walker has the chirality $|W_{(n-1)/2}\rangle$, then, the walker does not move.
The walker at time $t$ and position $x$ has a coin state expressed by the $n$-dimensional vector;
\[\Psi_{t}(x)={}^T\![\Psi_{t}^{W_{1}}(x), \Psi^{W_{2}}_{t}(x), \cdots, \Psi^{W_{n}}_{t}(x)]\in\mathbb{C}^{n}\]
and the time evolution is determined by the recurrence formula:
\begin{itemize}
\item if $n$ is odd, we have
\[\Psi_{t}(x)=V_{1}\Psi_{t-1}(x+(n-1)/2)+V_{2}\Psi_{t-1}(x+(n-3)/2)+\cdots+V_{n-1}\Psi_{t-1}(x-(n-3)/2)+V_{n}\Psi_{t-1}(x-(n-1)/2).\]
\item if $n$ is even, we have
\[\Psi_{t}(x)=V_{1}\Psi_{t-1}(x+n/2)+V_{2}\Psi_{t-1}(x+(n-2)/2)+\cdots+V_{n-1}\Psi_{t-1}(x-(n-2)/2)+V_{n}\Psi_{t-1}(x-n/2).\]
\end{itemize}
\noindent
Now let
\[\Psi_{t}= {}^T\!\left[\cdots,\begin{bmatrix}
\Psi_{t}^{W_{1}}(-1)\\
\vdots\\
\Psi_{t}^{W_{n}}(-1)\end{bmatrix},\begin{bmatrix}
\Psi_{t}^{W_{1}}(0)\\
\vdots\\
\Psi_{t}^{W_{n}}(0)\end{bmatrix},\begin{bmatrix}
\Psi_{t}^{W_{1}}(1)\\
\vdots\\
\Psi_{t}^{W_{n}}(1)\end{bmatrix},\cdots\right]\in(\mathbb{C}^{n})^{\mathbb{Z}}\]
and
\[U^{(s)}=SU_{x}=\begin{bmatrix}
\ddots&\vdots&\vdots&\vdots&\vdots&\cdots\\
\cdots&O&P_{-1}&O&O&O\cdots\\
\cdots&Q_{-2}&O&P_{0}&O&O\cdots\\
\cdots&O&Q_{-1}&O&P_{1}&O\cdots\\
\cdots&O&O&Q_{0}&O&P_{2}\cdots\\
\cdots&O&O&O&Q_{1}&O\cdots\\
\cdots&\vdots&\vdots&\vdots&\vdots&\ddots
\end{bmatrix}\;\;\;
with\;\;\;O=\begin{bmatrix}0&\cdots&0\\\vdots&\ddots&\vdots\\0&\cdots&0\end{bmatrix},\]
where $T$ means the transposed operation and $S$ is the standard shift operator:
\[S=\sum_{x}(|x\rangle\langle x+1|\otimes|L\rangle\langle L|+(|x\rangle\langle x-1|\otimes|R\rangle\langle R|).\]

The $\infty\times\infty$ unitary matrix $U^{(s)}$ is a time evolution operator on ${\cal H}$, that is, the state of the walker at position $x$ and time $t$ can be defined by
$\Psi_{t}=(U^{(s)})^{t}\Psi_{0}$.  \\
\indent
Now we prefer to introduce a map $\mu_{t}:\mathbb{Z}\to[0,1]$:
\[\mu_{t}(x)=\|\Psi_{t}(x)\|^{2}=|\Psi_{t}^{W_{1}}(x)|^{2}+\cdots+|\Psi^{W_{n}}_{t}(x)|^{2}\quad for\;\;x\in\mathbb{Z}.\]
Assuming $\|\Psi_{0}\|^{2}=1$, $\mu_{t}$ becomes the probability measure, and we can define the random variable $X_{t}$ from $\mu_{t}$, that is, the walker can be observed at position $x$ and time $t$ with the probability 
\[P(X_{t}=x)=\mu_{t}.\]
Here we define the stationary measure. Put $\mathbb{R}_{+}=[0,\infty)$, and we set a map 
$\phi:(\mathbb{C}^{n})^{\mathbb{Z}}\rightarrow \mathbb{R}_{+}^{\mathbb{Z}}$
such that for
\[\Psi= {}^T\!\left[\cdots,\begin{bmatrix}
\Psi^{W_{1}}(-1)\\
\vdots\\
\Psi^{W_{n}}(-1)\end{bmatrix},\begin{bmatrix}
\Psi^{W_{1}}(0)\\
\vdots\\
\Psi^{W_{n}}(0)\end{bmatrix},\begin{bmatrix}
\Psi^{W_{1}}(1)\\
\vdots\\
\Psi^{W_{n}}(1)\end{bmatrix},\cdots\right]\in(\mathbb{C}^{n})^{\mathbb{Z}},\]
we set the measure by
\[\mu:\mathbb{Z}\to\mathbb{R}_{+}\;such\;that\;\mu(x)=\phi(\Psi(x)) = |\Psi^{W_{1}}(x)|^{2} +\cdots+ |\Psi^{W_{n}}(x)|^{2}\;\;\;(x\in\mathbb{Z})\]
where $\Psi^{j}(x)\;(j=W_{1},\cdots,W_{n})$ is called {\it the stationary amplitude}.
Now let 
\begin{align}
\Sigma_{s}=\{\phi(\Psi_{0})\in\mathbb{R}_{+}^{\mathbb{Z}}: there\;exists\;\Psi_{0}
\;such\;that\;\;\phi((U^{(s)})^{t}\Psi_{0})=\phi(\Psi_{0})\;for\;any\;t\geq 0\},\end{align}
and we call the element of $\Sigma_{s}$, {\it the stationary measure} of the QW.
Now we consider the eigenvalue problem of the QW
\begin{eqnarray}
U^{(s)}\Psi^{(\lambda)}=\lambda\Psi^{(\lambda)}\quad(\lambda\in\mathbb{C}).\label{koyuti}
\end{eqnarray}
Since the unitarity of $U^{(s)}$, we have $|\lambda|=1$, and therefore we see $\phi(\Psi^{(\lambda)})\in{\cal M}_{s}$.

\section{Case 1: 2-state diagonal QW with one defect}
\label{onedefectqw}
In this section, we study the DTQW whose quantum coins are given by the set of diagonal unitary matrices with one defect at the origin:

\begin{align}\label{def1}\{U_{x}\}=\left\{
\left[ 
\begin{array}{cc}
e^{i\sigma_{\pm}} & 0 \\
0 & \bigtriangleup^{(\pm)} e^{-i\sigma_{\pm}}\\
\end{array} 
\right] _{ x=\pm1,\pm2,\cdots},
\left[ 
\begin{array}{cc}
a & b \\
c & d \\
\end{array} 
\right]_{x=0}\right\},\end{align}
where $\sigma_{\pm}\in[0,2\pi),\;a,b,c,d\in\mathbb{C},$ and $\bigtriangleup^{(\pm)}\in\mathbb{C}$ with $|\bigtriangleup^{(\pm)}|=1$. Note that Eq. \eqref{def1} is defined by the double sign.
\noindent
We assign the two different quantum coins to the positive and negative parts, respectively.
For the unitarity, we have $|a|^{2}+|c|^{2}=|b|^{2}+|d|^{2}=1,\;a\overline{b}+c\overline{d}=0,\;c=-\bigtriangleup\overline{b},$ and $d=\bigtriangleup\overline{a}$, where $\bigtriangleup\in\mathbb{C}$ is the determinant of $U_{0}$ with $|\bigtriangleup|^{2}=1$.
Here we show the stationary measure obtained by using the Splitted generating function method (the SGF method) developed in the previous studies \cite{watanabe, endosan, segawa}. The detail of the derivation of Theorem \ref{stat1} is given in Section \ref{proof1}.
\par\indent
\par\noindent
\begin{theo}\label{stat1}
The stationary measure of the QW is written by
\begin{align*}\mu(x)=\left\{ \begin{array}{ll}
(1+|b|^{2})|\alpha|^{2}+|a|^{2}|\beta|^{2}-2\Re{(\overline{a}b\overline{\alpha}\beta)} & (x\geq 1) \\
\\
|\alpha|^{2}+|\beta|^{2} & (x=0) \\
\\
|a|^{2}|\alpha|^{2}+(1+|b|^{2})|\beta|^{2}+2\Re(\overline{a}b\overline{\alpha}\beta)  & (x\leq-1) \\
\end{array} \right.,\end{align*}
\par\indent
\par\noindent
where $\alpha=\Psi^{L}(0), \;\beta=\Psi^{R}(0)$, and $\Re(x)$ is the real part of $x\;(x\in\mathbb{C})$.
Here the eigenvalues which give the stationary measure are $\lambda=\pm\sqrt{\bigtriangleup^{(\pm)}}$.
\end{theo}
\par\indent
\par\noindent
We see that the stationary measure does not depend on the two different quantum coins, that is, if $\sigma_{+}=\sigma_{-}$, then, we get the same result. 
We should remark that the stationary measure does not have an exponential decay for the position, which is in remarkable contrast to the QWs we treated in our previous studies \cite{watanabe,endosan}.
In addition, the stationary measure is generally non-uniform for the position, however, if $b=0$, that is, the defect is defined by the diagonal unitary matrix, or if 
\[|\alpha|^{2}-|\beta|^{2}-2\dfrac{\Re{(\overline{a}b\overline{\alpha}\beta)}}{|b|^{2}}=0,\]
then, the stationary measure becomes unitary measure. 
Here one of the interesting future problems is to elucidate the whole picture of the set of the stationary measure.
\par\noindent

\subsection{Proof of Theorem \ref{stat1}}
\label{proof1}
Taking advantage of the SGF method, let us solve the eigenvalue problem
\begin{align}U^{(s)}\Psi=\lambda\Psi,\label{eigenvalueproblem}\end{align}
where $\lambda\in\mathbb{C}$ with $|\lambda|=1$.
Rewriting the eigenvalue problem, we have
\begin{align}
\lambda\Psi(x)=P_{x+1}\Psi(x+1)+Q_{x-1}\Psi(x-1).
\label{bisyou}
\end{align}
Now Eq. \eqref{bisyou} can be expressed according to each position as follows:
\par\indent
\par\noindent
\begin{enumerate}
\item Case of $x=\pm2,\pm3,\cdots$.\\
\begin{eqnarray*}
\lambda\Psi^{L}(x)\!\!\!&=&\!\!\!e^{i\sigma_{\pm}}\Psi^{L}(x+1),\label{2phase-koyuuti-relation2.1}\\
\lambda\Psi^{R}(x)\!\!\!&=&\!\!\!\bigtriangleup^{(\pm)} e^{-i\sigma_{\pm}}\Psi^{R}(x-1).
\label{2phase-koyuuti-relation2.2}\end{eqnarray*}
\item Case of $x=1$.\\
\begin{eqnarray*}
\lambda\Psi^{L}(1)\!\!\!&=&\!\!\!e^{i\sigma_{+}}\Psi^{L}(2),\label{2phase-koyuuti-relation3.1}\\
\lambda\Psi^{R}(1)\!\!\!&=&\!\!\!c\Psi^{L}(0)+d\Psi^{R}(0).
\label{2phase-koyuuti-relation3.2}\end{eqnarray*}
\item Case of $x=0$.
\begin{eqnarray*}
\lambda\Psi^{L}(0)\!\!\!&=&\!\!\!e^{\sigma_{+}}\Psi^{L}(1),\label{2phase-koyuuti-relation5.1}\\
\lambda\Psi^{R}(0)\!\!\!&=&\!\!\!\bigtriangleup^{(-)} e^{-i\sigma_{-}}\Psi^{R}(-1).
\label{2phase-koyuuti-relation5.2}\end{eqnarray*}
\item Case of $x=-1$.\\
\begin{eqnarray*}
\lambda\Psi^{L}(-1)\!\!\!&=&\!\!\!a\Psi^{L}(0)+b\Psi^{R}(0),\label{2phase-koyuuti-relation4.1}\\
\lambda\Psi^{R}(-1)\!\!\!&=&\!\!\!\bigtriangleup^{(-)} e^{-i\sigma}\Psi^{R}(-2).
\label{2phase-koyuuti-relation4.2}\end{eqnarray*}

\end{enumerate}
Here we introduce the generating functions of $\Psi^{j}(x)\;(j=L,R)$ as follows:
\begin{align}
f^{j}_{+}(z)=\sum^{\infty}_{x=1}\Psi^{j}(x)z^{x},\quad
f^{j}_{-}(z)=\sum_{x=-1}^{-\infty}\Psi^{j}(x)z^{x}.
\label{2phase-bokansuu}\end{align}
Then we obtain
\par\indent
\par\noindent
\begin{lemma}
\label{2phase-hodai1}
Put
\begin{eqnarray*}A_{\pm}\!\!\!&=&\!\!\!\begin{bmatrix}\lambda-\dfrac{e^{i\sigma_{\pm}}}{z}&0 \nonumber\\\
0&\lambda-\bigtriangleup^{(\pm)} e^{-i\sigma_{\pm}}z
\end{bmatrix},
\;{\bf f}_{\pm}(z)=\left[\begin{array}{c}f^{L}_{\pm}(z)\\f^{R}_{\pm}(z)\end{array}\right],\nonumber\\\
{\bf a}_{+}(z)\!\!\!&=&\!\!\!\left[\begin{array}{c}-\lambda\alpha\\ (c\alpha+d\beta)z\end{array}\right],\;
{\bf a}_{-}(z)=\left[\begin{array}{c}(a\alpha+b\beta)z^{-1}\\-\lambda\beta\end{array}\right].\end{eqnarray*}
\par\indent
\par\noindent
Then, 
\begin{align}
A_{\pm}{\bf f}_{\pm}(z)={\bf a}_{\pm}(z)\label{2phase-hodai1-houteisiki}
\end{align}
hold.
\end{lemma}
\par\indent
\par\noindent
Noting
\begin{align}
\det A_{\pm}=-\dfrac{\lambda\bigtriangleup^{(\pm)} }{ e^{i\sigma_{\pm}}z}\left\{z^{2}-\dfrac{e^{i\sigma_{\pm}}}{\lambda\bigtriangleup^{(\pm)}}(\lambda^{2}+\bigtriangleup^{(\pm)})z+\dfrac{e^{2i\sigma_{\pm}}}{\bigtriangleup^{(\pm)}}\right\},
\label{deta01sou}
\end{align}
we chose $\theta_{s}^{(\pm)}, \> \theta_{l}^{(\pm)} \in\mathbb{C}$ satisfying
\begin{align}
\det A_{\pm}=-\dfrac{\lambda\bigtriangleup^{(\pm)} }{ e^{i\sigma_{\pm}}z}(z-\theta_{s}^{(\pm)})(z-\theta_{l}^{(\pm)})
\label{deta02sou}
\end{align}
and $|\theta_{s}^{(\pm)}|\leq1\leq|\theta_{l}^{(\pm)}|$.
Here Eqs. \eqref{deta01sou} and \eqref{deta02sou} give $\theta_{s}^{(\pm)}\theta_{l}^{(\pm)}=(\bigtriangleup^{(\pm)})^{-1} e^{2i\sigma_{\pm}}$.\\
\\
From now on, let us derive $f_{\pm}^{L}(z)$ and $f_{\pm}^{R}(z)$ by solving Eq. \eqref{2phase-hodai1-houteisiki} in Lemma \ref{2phase-hodai1}.
\par\indent
\par\noindent
\begin{enumerate}
\item Case of $f_{+}^{L}(z)$. Eq. (\ref{2phase-hodai1-houteisiki}) gives
\begin{align*}
f^{L}_{+}(z)
=\dfrac{\lambda\bigtriangleup^{(+)}\alpha }{e^{i\sigma_{+}} \det A} \left\{z-\dfrac{\lambda e^{i\sigma_{+}}}{\bigtriangleup^{(+)}}\right\}.
\end{align*}
Putting $\theta_{s}=\lambda(\bigtriangleup^{(+)})^{-1}e^{i\sigma_{+}}$, we have
\begin{align*}
f^{L}_{+}(z)
&=\dfrac{\alpha\theta_{l}^{-1}z}{1-(\theta_{l}^{(+)})^{-1}z}
\\
&=\alpha(\theta_{l}^{(+)})^{-1}z+\alpha(\theta_{l}^{(+)})^{-2}z^{2}+\cdots.
\end{align*}
Hence we see 
\begin{align}
f_{+}^{L}(z)=\alpha\sum^{\infty}_{x=1}(\theta_{l}^{(+)})^{-x}z^{x}=\alpha\sum^{\infty}_{x=1}(\bigtriangleup^{(+)} e^{-2i\sigma_{+}})^{x}(\theta_{s}^{(+)})^{x}z^{x}.
\label{panda01sou}
\end{align}
Equation \eqref{panda01sou} and the definition of $f_{+}^{L}(z)$ imply
\begin{align}
\Psi^{L}(x)=\alpha(\bigtriangleup^{(+)} e^{-2i\sigma_{+}})^{x}(\theta_{s}^{(+)})^{x}\;\;\;(x=1,2,\cdots),
\end{align}
where
\begin{align}
\theta_{s}^{(+)}=\lambda(\bigtriangleup^{(+)})^{-1}e^{i\sigma_{+}}.
\label{onene01sou}
\end{align}
\item Case of $f_{+}^{R}(z)$. 
Putting $\theta_{s}^{(+)}=\lambda^{-1}e^{i\sigma_{+}}$, we have from Eq. (\ref{2phase-hodai1-houteisiki})
\begin{align}
f^{R}_{+}(z)
&=- \dfrac{(\bigtriangleup^{(+)})^{-1}e^{i\sigma_{+}}(c\alpha+d\beta)z}{z-\theta_{l}^{(+)}}
\nonumber
\\
&=\dfrac{(\bigtriangleup^{(+)})^{-1}e^{i\sigma_{+}}(c\alpha+d\beta)(\theta_{l}^{(+)})^{-1}z}{1-(\theta_{l}^{(+)})^{-1}z}\nonumber\\
&=(\bigtriangleup^{(+)})^{-1}e^{i\sigma_{+}}(c\alpha+d\beta)\sum_{x=1}^{\infty}((\theta_{l}^{(+)})^{-1}z)^{x}\nonumber\\
&=(\bigtriangleup^{(+)})^{-1}e^{i\sigma_{+}}(c\alpha+d\beta)\sum_{x=1}^{\infty}(\bigtriangleup^{(+)} e^{-2i\sigma_{+}}\theta_{s}^{(+)}z)^{x}.
\label{panda02sou}
\end{align}
Equation \eqref{panda02sou} and the definition of $f_{+}^{R}(z)$ give
\begin{align}
\Psi^{R}(x)=(\bigtriangleup^{(+)})^{-1}e^{i\sigma_{+}}(c\alpha+d\beta)(\bigtriangleup^{(+)} e^{-2i\sigma_{+}}\theta_{s}^{(+)})^{x}\;\;\;(x=1,2,\cdots),
\end{align}
where
\begin{align}
\theta_{s}^{(+)} = \lambda^{-1}e^{i\sigma_{+}}.
\label{onene02sou}
\end{align}
\item Case of $f_{-}^{L}(z)$.
Putting $(\theta_{l}^{(-)})^{-1} = \lambda^{-1}\bigtriangleup^{(-)} e^{-i\sigma_{-}}$,
Eq. \eqref{2phase-hodai1-houteisiki} gives
\begin{align*}
f^{L}_{-}(z)
&= -\dfrac{(\bigtriangleup^{(-)})^{-1}e^{i\sigma_{-}}(a\alpha+b\beta)(\theta_{s}^{(-)})^{-1}(\theta_{l}^{(-)})^{-1}z^{-1}}{z^{-1}-(\theta_{s}^{(-)})^{-1}}
.
\end{align*}
Hence we have
\begin{align}
f^{L}_{-}(z)
&=\dfrac{(\bigtriangleup^{(-)})^{-1}e^{i\sigma_{-}}(a\alpha+b\beta)(\theta_{l}^{(-)})^{-1}z^{-1}}{1-\theta_{s}^{(-)}z^{-1}}\\
&=\dfrac{e^{-i\sigma_{-}}(a\alpha+b\beta)\theta_{s}^{(-)}z^{-1}}{1-\theta_{s}^{(-)}z^{-1}}\\
&=e^{-i\sigma_{-}}(a\alpha+b\beta)\sum^{-\infty}_{x=-1}(\theta_{s}^{(-)})^{-x}z^{x}.
\label{panda03sou}
\end{align}
Equation \eqref{panda03sou} and the definition of $f^{L}_{-}(z)$ yield
\begin{align*}
\Psi^{L}(x)=e^{-i\sigma_{-}}(a\alpha+b\beta)(\theta_{s}^{(-)})^{-x}\;\;\;(x=-1,-2,\cdots),
\end{align*}
where
\begin{align}
\theta_{s}^{(-)}= \lambda^{-1}e^{i\sigma_{-}}.
\label{onene03sou}
\end{align}
\item Case of $f_{-}^{R}(z)$. Putting $
(\theta_{l}^{(-)})^{-1}= \lambda e^{-i\sigma_{-}},
$
we have from Eq. \eqref{2phase-hodai1-houteisiki} 

\begin{align}
f^{R}_{-}(z)=\beta \sum^{-\infty}_{x=-1}((\theta_{s}^{(-)})^{-1}z)^{x}.
\label{panda04sou}
\end{align}
Therefore, Eq. \eqref{panda04sou} and the definition of  $f^{R}_{-}(z)$ give
\begin{align*}
\Psi^{R}(x)=\beta (\theta_{s}^{(-)})^{-x}\;\;\;(x=-1,-2,\cdots),
\end{align*}
where
\begin{align}
\theta_{s}^{(-)}=\lambda (\bigtriangleup^{(-)})^{-1}e^{i\sigma_{-}}.
\label{onene04sou}
\end{align}
\end{enumerate} 
As a result, we obtain

\begin{align}\Psi(x)=\left\{ \begin{array}{ll}
\begin{bmatrix}\alpha(\bigtriangleup^{(+)}e^{-2i\sigma_{+}}\theta^{(+)}_{s})^{x}\\
(\bigtriangleup^{(+)})^{-1}(c\alpha+d\beta)e^{i\sigma_{+}}(\bigtriangleup^{(+)}e^{-2i\sigma_{+}}\theta^{(+)}_{s})^{x} \end{bmatrix}& (x\geq 1) \\
\\
\begin{bmatrix}\alpha\\ 
\beta\end{bmatrix} & (x=0) \\
\\
\begin{bmatrix}(a\alpha+b\beta)e^{-i\sigma_{-}}(\theta^{(-)}_{s})^{|x|}\\ 
\beta (\theta^{(-)}_{s})^{|x|} \end{bmatrix}& (x\leq-1) \\
\end{array} \right..\label{araisan2sou}\end{align}
\par\indent
\par\noindent
Moreover, $4$ expressions of $\theta_{s}$, that is, Eqs. \eqref{onene01sou}, \eqref{onene02sou},
\eqref{onene03sou}, and \eqref{onene04sou} suggest
\begin{eqnarray*}
\theta_{s}^{(+)}=\lambda(\bigtriangleup^{(+)})^{-1} e^{i\sigma_{+}}=\lambda^{-1}e^{i\sigma_{+}}
\end{eqnarray*}
and
\begin{eqnarray*}
\theta_{s}^{(-)}=\lambda^{-1} e^{i\sigma_{-}}=\lambda(\bigtriangleup^{(-)})^{-1} e^{i\sigma_{-}}.
\end{eqnarray*}
Therefore, we see $\lambda=\pm\sqrt{\bigtriangleup^{(\pm)}}$ and $|\theta^{(\pm)}_{s}|=1$, by choosing the sign in a suitable way.
Now the proof of Theorem \ref{stat1} is completed.

\par\indent
\par\noindent

\section{Case 2: $3$-state case.}
\label{3state-1defect}
\subsection{$3$-state diagonal QW with one defect}
In this subsection, we study the DTQW whose quantum coins are given by the set of diagonal unitary matrices with one defect at the origin:

\begin{align}\label{def2}U_{x}=\left\{
\left[ 
\begin{array}{ccc}
e^{i\sigma_{\pm}} & 0&0 \\
0 & e^{-i\sigma_{\pm}}&0\\
0&0&\bigtriangleup^{(\pm)} 
\end{array} 
\right]_{x=\pm1,\pm2,\cdots},
\left[ 
\begin{array}{ccc}
a & b &c\\
d & e& f \\
g &h&i
\end{array} 
\right]_{x=0}\right\},\end{align}
where $\sigma_{\pm}\in[0,2\pi),\;a,b,c,d,e,f,g,h,i\in\mathbb{C},$ and $\bigtriangleup^{(\pm)}\in\mathbb{C}$ with $|\bigtriangleup^{(\pm)}|=1$. Here Eq. \eqref{def2} is defined by the double sign.
\noindent
We give the two different quantum coins in the positive and negative parts, respectively.
Here we show the stationary measure obtained by using the SGF method \cite{watanabe, endosan,segawa}. 
\par\indent
\par\noindent
\begin{theo}\label{stat3}
The stationary measure of the QW is written by

\begin{align*}\mu(x)=\left\{ \begin{array}{ll}
(1+|g|^{2})|\alpha|^{2}+|h|^{2}|\gamma|^{2}+|i|^{2}|\beta|^{2}+2\Re{(g\alpha\overline{h}\overline{\gamma}+g\alpha\overline{i}\overline{\beta}+g\gamma\overline{i}\overline{\beta})}& (x\geq 1) \\
\\
|\alpha|^{2}+|\gamma|^{2}+|\beta|^{2} & (x=0) \\
\\
|\alpha|^{2}+(1+|c|^{2})|\beta|^{2}+|b|^{2}|\gamma|^{2}+2\Re(a\alpha\overline{b}\overline{\gamma}+a\alpha\overline{c}\overline{\beta}+b\gamma\overline{c}\overline{\beta})& (x\leq-1) \\
\end{array} \right.,\end{align*}
\par\indent
\par\noindent
where $\alpha=\Psi^{L}(0), \;\beta=\Psi^{R}(0),\;\xi=\Psi^{L}(1),\;\gamma=\Psi^{R}(-1)$, and $\Re(x)$ means the real part of $x\;(x\in\mathbb{C})$.\\
Note that the eigenvalues which contribute to the stationary measure are $\lambda=\pm\sqrt{\bigtriangleup^{(\pm)}e^{i\sigma_{\pm}}}$.

\end{theo}
\par\indent
\par\noindent
The stationary measure does not depend on the two different quantum coins, that is, if $\sigma_{+}=\sigma_{-}$, then, we have the same measure. 
We should also remark that the stationary measure does not have an exponential decay for the position.
Furthermore, the defect and initial coin state strongly influence on the stationary measure.
In addition, the stationary measure is generally non-uniform for the position, however, if the defect is defined by diagonal unitary matrix, then the stationary measure becomes a uniform measure. 

\subsection{$3$-state diagonal QW}
\label{3state-homogeneous}
Here we consider the $3$-state space-homogeneous QW, whose time-evolution is described by the following diagonal unitary matrix:\\
\begin{align}
\label{def2}
U=\begin{bmatrix}
e^{i\sigma}&0&0\\
0&e^{-i\sigma}&0\\
0&0&\bigtriangleup
\end{bmatrix},
\end{align}
where $\bigtriangleup\in\mathbb{C}$ is the determinant of $U$ with $|\bigtriangleup|=1$.\\
\indent
This is a special case of the 3-state diagonal QW with one defect. Now we show the stationary measure. We obtain Theorem \ref{stat9} in a similar way as Subsection \ref{proof1}, and we omit the proof here.
\par\indent
\par\noindent
\begin{theo}
\label{stat9}
The stationary measure of the QW is written by
\[\mu(x)=\left\{ \begin{array}{ll}
|\alpha|^{2}+|\beta|^{2} & (x\neq0) \\
|\alpha|^{2}+|\gamma|^{2}+|\beta|^{2} & (x=0) \\
\end{array} \right.,\]
where $\alpha=\Psi^{L}(0), \;\beta=\Psi^{R}(0),$ and $\gamma=\Psi^{O}(0)$.
Remark that the eigenvalues are $\lambda=\pm\sqrt{\bigtriangleup e^{i\sigma}}$.

\end{theo}
\par\indent
\par\noindent
By putting $a=e^{i\sigma},\;e=e^{-i\sigma},\;i=\bigtriangleup$, we also obtain the same outcome from Theorem \ref{stat3}.
We emphasize that the stationary measure depends on position $x$, which is in marked contrast to that of $2$-state homogeneous case \cite{takei}. One of our basic future problems is to obtain the description of all the stationary measure for the $3$-state homogeneous QW by applying the method developed in Ref. \cite{takei}.

\section{summary}
By using the SGF method \cite{watanabe, endosan, segawa}, we obtained the stationary measure for $3$-state homogeneous diagonal QW, $2$ and $3$-state diagonal QWs with one defect, and then, clarified the characteristic properties. Through all the cases we saw, we also found that the eigenvalues are strongly influenced by the determinant of the diagonal unitary matrices. From a viewpoint of classification, one of our essential future problems is to apply the method \cite{takei} to the diagonal QWs we treated, and make clear the whole description of the stationary measure. For instance, it is substantive to elucidate the conditions of the stationary measure to be homogeneous, which may lead to classify the diagonal QWs with defects. As our results suggest, it is precious to investigate the influence of the number of states and the diagonality of the defects on the uniformity of stationary measure. Also, to develop the method to construct the stationary measure for the QWs with plural defects is imperative. On the other side, it is interesting to consider the topological invariants for the 3-state cases, which may lead to establish the interpretation for localization of QWs from a viewpoint of condensed matter physics \cite{eko}. 

\setcounter{footnote}{0}
\renewcommand{\thefootnote}{\alph{footnote}}
\section*{Acknowledgement}
This work is supported by Toshiko Yuasa memorial fund, and the Grant-in-Aid for Scientific Research (Challenging Exploratory Research) of Japan Society for the Promotion of Science (Grant No.15K13443). \\
\par\indent
\par\noindent
\nonumsection{References}

\end{document}